# Fourier Analysis and Benford Random Variables

Frank Benford
June, 2020; revised November, 2020

## 1. Introduction

To forestall potential confusion, let me say at the outset that the author of this paper is not the physicist Frank Benford for whom Benford's law is named, but one of his grandsons.

This paper has several major purposes. Perhaps the central purpose is to summarize some results from investigations into base dependence of Benford random variables, including a discussion of Whittaker's astonishing random variable. The principal tools used to derive these results are Fourier series and Fourier transforms, and a second major purpose of this paper is to present an introductory exposition about these tools. Since there exist many excellent introductions to Fourier series, and a fair number of undergraduate level introductions to Fourier transforms, one may wonder if another introduction is really needed. My motivation for writing this paper is twofold. First, I believe that the theory of Benford random variables and the "Benford analysis" of a positive random variable has some interest and deserves to be better known. Second, I think Benford analysis provides a really excellent illustration of the utility of Fourier series and transforms, and reveals certain interconnections between series and transforms that are not obvious from the usual way these subjects are introduced.

**A note to student readers**. This paper in neither a research paper nor a survey of the literature. Instead, it's intended as an undergraduate level introduction to both Benford random variables and Fourier analysis. Consequently, the reader may have an imperfect understanding of some of the language. For example, you may not understand every detail in the statement "The space $L^2[0, 1]$ is a Hilbert space." If you don't understand everything in this sentence, my advice is not to worry about it. You should be able to get the gist of the ideas presented here without having a deep understanding of all of the language. If the language piques your interest, ask one of your professors (who may suggest that you'll want to enroll in a course in real analysis).

## 2. Benford Random Variables

My grandfather [1] studied the empirical distribution of "first digits" (i.e., the leftmost digit, or most significant digit) of randomly selected numbers by compiling a large (for 1938) and heterogeneous data set of 20,229 (decimal) numbers culled from newspapers, almanacs, scientific handbooks, and other sources. He found[1] that the proportion of these numbers with first digit $d$ (for every $d \in \{1, 2, \ldots, 9\}$) was given to a close approximation by the formula

---

[1] This was actually a rediscovery, as Simon Newcomb had made the same observation in 1881.

$$\Pr(D_1 = d) = \log_{10}\left(1 + \frac{1}{d}\right), \tag{2.1}$$

where $\Pr()$, for the moment, denotes "proportion." This formula is "The First Digit Law." The observed and expected proportions of each of the nine possible first digits (from grandfather's data set and eq. (2.1), respectively) are shown in the following figure.

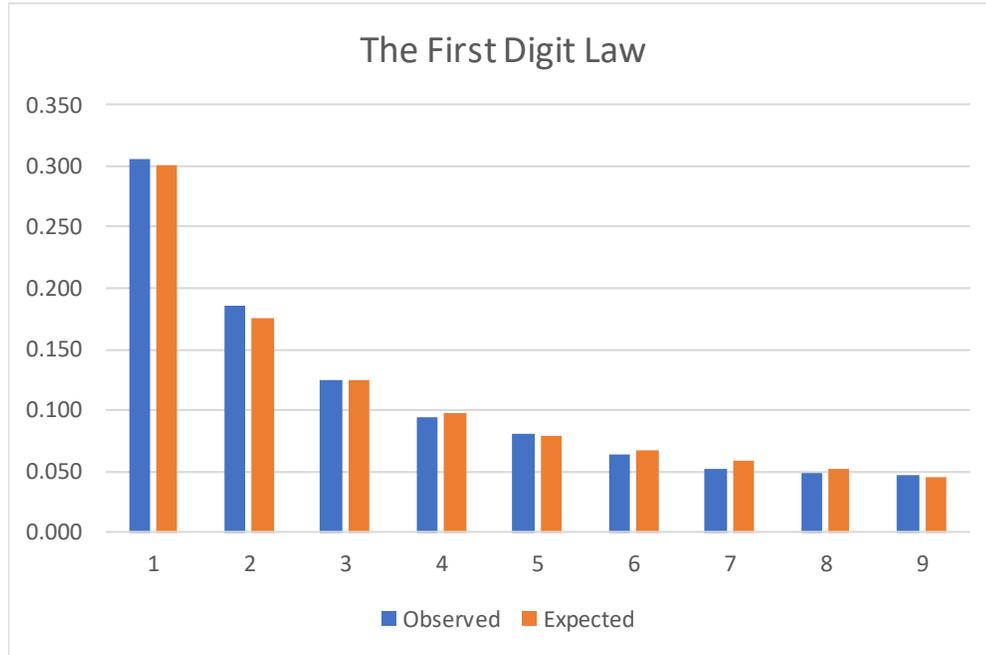

No finite data set can satisfy eq. (2.1) exactly, as the proportion of numbers in a data set with first digit $d$ is necessarily a rational number, whereas $\log_{10}(1 + 1/d)$ is irrational. However, we may conceptualize a "Benford" data set of size $n$ as being a random sample, i.e. $n$ independent realizations of a random variable $X$, such that eq. (2.1) holds with $\Pr()$ interpreted as "probability." This is one of the properties of a Benford random variable.

The First Digit Law is the most famous part of Benford's Law, but it's not the whole law. Benford's Law may be phrased in terms of the joint distribution of the leftmost $m$ digits of the numbers in a data set for all $m \in \mathbb{N}$, but it is better to phrase it in terms of the "significand" function. To explain this, consider two famous numbers from physics, written in "scientific notation." Let $c \equiv 2.998 \times 10^8$ meters per second, and $h \equiv 6.626 \times 10^{-34}$ Joule seconds. (These are approximations to the speed of light in vacuum and Planck's constant, respectively.) The two numbers 2.998 and 6.626 are the *significands* of $c$ and $h$, and we write $S_{10}(c) = 2.998$ and $S_{10}(h) = 6.626$. In general, any $x > 0$ may be written in the form

$$x = t \times 10^k \quad \text{with} \quad t \in [1, 10) \quad \text{and} \quad k \in \mathbb{Z},$$

and both $t$ and $k$ in this representation of $x$ are unique. The number $t$ in this representation of $x$ is the (base 10) "significand of $x$," and we write $S_{10}(x) \equiv t$. Significands in base 10 are most familiar to us, but other bases are of interest. Let $b > 1$. Any number $x > 0$ may be written in the form

$$x = t \times b^k \quad \text{with} \quad t \in [1, b) \quad \text{and} \quad k \in \mathbb{Z}. \tag{2.2}$$

Both $t$ and $k$ in this representation are unique, and the "base $b$ significand of $x$," written $S_b(x)$, is defined as this $t$. Hence,

$$x = S_b(x) \times b^k \quad \text{with} \quad S_b(x) \in [1, b) \quad \text{and} \quad k \in \mathbb{Z}. \tag{2.3}$$

Now let $X$ be a positive random variable; that is, $\Pr(X > 0) = 1$. Assume that $X$ is continuous with a probability density function (pdf). **Definition 2.1**: $X$ is base $b$ Benford (or $X$ is $b$-Benford) if and only if the cumulative distribution function (cdf) of $S_b(X)$ is given by

$$\Pr(S_b(X) \leq s) = \log_b(s) \quad \text{for all} \quad s \in [1, b). \tag{2.4}$$

It's easy to show that a base 10 Benford random variable satisfies the First Digit Law given by eq. (2.1). Let $D_1(X)$ denote the first (most significant) digit in the decimal representation of $X$, and observe that $D_1(X) = d$ iff $d \leq S_{10}(X) < d + 1$ for every $d \in \{1, 2, \ldots, 9\}$.

It's useful at this point to introduce some non-standard notation. Let $y \in \mathbb{R}$ and recall that the "floor" of $y$, written $\lfloor y \rfloor$, is defined as the largest *integer* that is less than or equal to $y$. Define $\langle y \rangle$ as:

$$\langle y \rangle \equiv y - \lfloor y \rfloor \tag{2.5}$$

and note that $0 \leq \langle y \rangle < 1$ for every $y \in \mathbb{R}$. We will call $\langle y \rangle$ the *fractional part* of $y$, though if $y < 0$ this description is misleading. For example, $\lfloor 3.14 \rfloor = 3$, so $\langle 3.14 \rangle = 0.14$, but $\lfloor -3.14 \rfloor = -4$, so $\langle -3.14 \rangle = -3.14 - (-4) = 0.86$.

If we take the logarithm base $b$ of eq. (2.3) we obtain

$$\log_b(x) = \log_b(S_b(x)) + k. \tag{2.6}$$

On the other hand,

$$\log_b(x) = \langle \log_b(x) \rangle + \lfloor \log_b(x) \rfloor. \tag{2.7}$$

As $\lfloor \log_b(x) \rfloor$ is necessarily an integer and $0 \leq \log_b(S_b(x)) < 1$, comparison of eqs. (2.6) and (2.7) shows that

$$\log_b(S_b(x)) = \langle \log_b(x) \rangle \quad \text{and} \quad k = \lfloor \log_b(x) \rfloor \tag{2.8}$$

for any $x > 0$. Using this notation, we may rephrase Definition 2.1 in several logically equivalent ways: $X$ is $b$-Benford iff

(1) $\Pr(\log_b(S_b(X)) \leq \log_b(s)) = \log_b(s) \quad$ for every $\quad s \in [1, b)$,

(2) $\Pr(\langle \log_b(X) \rangle \leq u) = u \quad$ for every $u \in [0, 1)$,

(3) $\langle \log_b(X) \rangle \sim U[0, 1)$,

(4) $X = b^Y$ where $\langle Y \rangle \sim U[0, 1)$,

where the notation "$W \sim U[0,1)$" means[2] that $W$ is uniformly distributed on the half open interval $[0,1)$.

For any random variable $Y$, if $\langle Y \rangle \sim U[0,1)$ we sometimes say that $Y$ is "uniformly distributed modulo one," abbreviated "u. d. mod 1." Hence $X$ is $b$-Benford iff $\log_b(X)$ is u. d. mod 1.

Given a positive random variable $X$ and $b > 1$, we may ask (1) whether or not $X$ is $b$-Benford; and if it's not, (2) to what extent does it deviate from "Benfordness." I call an investigation of this sort a "Benford analysis." Let's establish some nomenclature for a Benford analysis. First, define

$$Y \equiv \log_b(X) = \Lambda_b \ln(X) \quad \text{where} \quad \Lambda_b \equiv \frac{1}{\ln(b)}. \tag{2.9}$$

Next, let

$$G() \text{ and } g() \quad \text{denote the cdf and pdf of } Y,$$
$$\widetilde{G}() \text{ and } \widetilde{g}() \quad \text{denote the cdf and pdf of } \langle Y \rangle.$$

Given $\widetilde{g}()$ we may answer the two questions given above. (1) $X$ is $b$-Benford iff $\widetilde{g}(u) = 1$ for almost all $u \in [0,1)$. (2) If $X$ is not $b$-Benford, we may measure its deviation from Benfordness by any measure of the deviation of $\widetilde{g}()$ from a uniform distribution. For example, if $\widetilde{g}()$ is continuous, or if its only discontinuities are "jumps," we could use the infinity norm:

$$\|\widetilde{g} - 1\|_\infty \equiv \sup(|\widetilde{g}(u) - 1|: 0 \le u < 1). \tag{2.10}$$

We need a way to find $\widetilde{g}()$ from $g()$. Under a reasonable assumption, it may be shown that

$$\widetilde{g}(u) = \sum_{k \in \mathbb{Z}} g(k + u) \tag{2.11}$$

for all $u \in [0,1)$. To see this, and to explain the "reasonable assumption," begin with the cdf of $\langle Y \rangle$. For any $u \in [0,1)$,

$$\widetilde{G}(u) = \Pr(\langle Y \rangle \le u) = \sum_{k \in \mathbb{Z}} \Pr(k \le Y \le k + u)$$
$$= \sum_{k \in \mathbb{Z}} [G(k + u) - G(k)]. \tag{2.12}$$

If $Y$ has finite support, we may replace the series in eq. (2.12) by a finite sum

$$\widetilde{G}(u) = \sum_{k=-N}^{N} \left[ G(k+u) - G(k) \right]$$

for some $N \in \mathbb{N}$, and term by term differentiation yields

$$\widetilde{g}(u) = \sum_{k=-N}^{N} g(k+u). \tag{2.13}$$

Similarly, simple term by term differentiation of eq. (2.12) yields eq. (2.11). The problem with this "proof" is that term by term differentiation of an infinite series is not always valid. The condition that's sufficient to justify this procedure is that the series on the right-hand side of eq. (2.11) converge *uniformly* (Rudin [7], Theorem 7.17). While it's certainly possible to impose weak conditions on $g()$ that imply the uniform convergence of the series in eq. (2.11), these conditions are somewhat artificial, and (in my opinion) it's better just to assume that this condition holds.

Although eq. (2.11) is fundamental for a Benford analysis of $X$, it is not very useful for finding the answers to some analytical questions one may ask. For example, suppose that $X \sim \text{Lognormal}(\mu, \sigma^2)$, so $\log_b(X) = \Lambda_b \ln(X) \sim N\left( \Lambda_b \mu, (\Lambda_b \sigma)^2 \right)$. As we'll show in Section 4, $X$ is not $b$-Benford, but the deviation $\| \widetilde{g} - 1 \|_\infty$ is generally quite small, so a Lognormal random variable is almost $b$-Benford. One may wish to investigate how this deviation varies as a function of $\mu$ and $\sigma^2$. For a second example, suppose that $X$ is $b$-Benford. We may ask: what is the set $\{ c > 1 \colon X \text{ is } c\text{-Benford} \}$? This set certainly includes $b$, but whether it includes any $c \neq b$ is unclear at this point. Equation (2.11) does not seem very useful for an investigation into these questions.

Fourier analysis provides the tools needed to answer questions like these. The Fourier series representation of $\widetilde{g}(u)$ is[3]

$$\widetilde{g}(u) = \sum_{n \in \mathbb{Z}} c_n e^{2\pi i n u} \tag{2.14}$$

where the Fourier coefficients $c_n$ for all $n \in \mathbb{Z}$ are given by

$$c_n \equiv \int_0^1 e^{-2\pi i n u} \, \widetilde{g}(u) \, du. \tag{2.15}$$

At first sight, these expressions must look very unpromising for two reasons, one superficial and the second significant. The superficial reason is that the complex exponentials

$$e^{2\pi i n u} = \cos(2\pi n u) + i \sin(2\pi n u)$$

are complex-valued, so the series of real-valued functions of eq. (2.11) has been rewritten as a series of complex-valued functions multiplied by complex coefficients. This may seem like one step forward and two steps back, but as we'll see in the next section, eq. (2.14) can always be rewritten as a series of real-valued functions multiplied by real-valued coefficients. The significant reason is that computation of the Fourier coefficients given by eq. (2.15) apparently requires that we already know $\widetilde{g}()$, the function for which we're trying to find a useful expression. As it happens, Fourier transforms come to our rescue here. We may show that

$$c_n = \widehat{g}(n) \tag{2.16}$$

where $\widehat{g}()$ denotes the Fourier transform of $g()$. In words, we may obtain the Fourier *coefficients* of $\widetilde{g}()$ from the Fourier *transform* of $g()$. This is the key result that allows our Fourier analysis to proceed. All one needs to know to understand the following proof is the definition of "the Fourier transform of $g()$": for any $\xi \in \mathbb{R}$,

$$\widehat{g}(\xi) \equiv \int_{-\infty}^{\infty} e^{-2\pi i \xi y} \, g(y) \, dy. \tag{2.17}$$

---

[3] The sense of the equality in eq. (2.14) is explained in Section 3.

Also, this proof makes implicit use of the uniform convergence of the series in eq. (2.11) (Rudin [7], Theorem 7.16).

**Proof** (of eq. (2.16)).

$$\begin{aligned} c_n &= \int_0^1 e^{-2\pi i n u} \, \widetilde{g}(u) \, du = \int_0^1 e^{-2\pi i n u} \sum_{k \in \mathbb{Z}} g(k+u) \, du \\ &= \sum_{k \in \mathbb{Z}} \int_0^1 e^{-2\pi i n(k+u)} g(k+u) \, du = \int_{-\infty}^{\infty} e^{-2\pi i n y} g(y) \, dy = \widehat{g}(n), \end{aligned}$$

as was to be shown.

Combining eqs. (2.11), (2.14), and (2.16), we obtain

$$\sum_{k \in \mathbb{Z}} g(k+u) = \sum_{n \in \mathbb{Z}} \widehat{g}(n) e^{2\pi i n u}. \tag{2.18}$$

This equation is known as the Poisson summation formula (Stein and Shakarchi [9], page 154).

We pause, then, in our exposition about Benford analyses, to list the essential facts about Fourier series and Fourier transforms that are required to continue the argument. The reader who is familiar with Fourier series and Fourier transforms may skip directly to Section 4.

## 3. Fourier Series and Fourier Transforms

Elements of $L^2[0, 1]$ are functions from $[0, 1]$ into $\mathbb{C}$, but we will mainly be concerned with real valued functions. The space $L^2[0, 1]$ is a Hilbert space, with an *inner product* defined as

$$(f, g) \equiv \int_0^1 f(u)\overline{g(u)}\, du \tag{3.1}$$

for any $f$ and $g$ in $L^2[0, 1]$, where the overbar denotes complex conjugation. This inner product defines a *norm*[4]: for any $f \in L^2[0, 1]$,

$$\|f\| \equiv \sqrt{(f, f)} = \left[ \int_0^1 |f(u)|^2 \right]^{1/2}. \tag{3.2}$$

It is known that the collection of complex exponentials $(\phi_n(): n \in \mathbb{Z})$ where

$$\phi_n(u) \equiv e^{2\pi i n u} \tag{3.3}$$

forms a complete orthonormal basis for $L^2[0, 1]$. That is, for any $f \in L^2[0, 1]$ there exists a doubly infinite sequence of (generally complex) scalars $(c_n: n \in \mathbb{Z})$ such that

$$f(u) = \sum_{n \in \mathbb{Z}} c_n \phi_n(u) = \sum_{n \in \mathbb{Z}} c_n e^{2\pi i n u} \tag{3.4}$$

where equality in eq. (3.4) is "in the sense of the norm," i.e.,

$$\lim_{N \to \infty} \left\| f() - \sum_{n=-N}^{N} c_n \phi_n() \right\| = 0. \tag{3.5}$$

(There may be points $u \in [0, 1]$ where equality in eq. (3.4) doesn't hold. For example, it will not generally hold at points $u$ where $f$ is not continuous. Equation (3.5), however, implies that the set of points $u$ where equality does not hold has measure zero[5].)

The expression on the right-hand side of eq. (3.4) is the "Fourier series expansion" for $f$. The scalars $(c_n: n \in \mathbb{Z})$ are the "Fourier coefficients" of $f$, and are given by

$$c_n = (f, \phi_n) = \int_0^1 e^{-2\pi i n u} f(u)\, du. \tag{3.6}$$

Now assume that $f$ is real valued. Equations (3.4) and (3.6) state that $f()$ can be written as a series whose terms are complex valued functions $\phi_n()$ multiplied by complex scalars $c_n$. Complex exponentials are algebraically convenient, but using them imposes the cost of working with complex functions and complex coefficients. However, it may be shown that eq. (3.4) can always be rewritten as a series with real valued functions and real coefficients. To be specific,

---

[4] Technically, eq. (3.2) is a only a "seminorm." The usual way to get around this difficulty is to redefine equality for elements of $L^2[0, 1]$. If $f$ and $g$ are elements of $L^2[0, 1]$, we say $f$ and $g$ are *equal* and write $f = g$ iff $\{x \in [0, 1]: f(x) \neq g(x)\}$ has measure zero.

[5] This innocuous sounding statement is actually a nontrivial theorem that was first proven by Lennart Carleson in 1966.

$$f(u) = \frac{1}{2}a_0 + \sum_{n=1}^{\infty}[a_n\cos(2\pi nu) + b_n\sin(2\pi nu)] \tag{3.7}$$

where

$$
\begin{aligned}
a_n &\equiv 2\int_0^1 \cos(2\pi nu)f(u)\,du, && n \in \{0\} \cup \mathbb{N}, \\
b_n &\equiv 2\int_0^1 \sin(2\pi nu)f(u)\,du, && n \in \mathbb{N}.
\end{aligned}
\tag{3.8}
$$

To prove these formulas, first note that $c_0 = \int_0^1 f(u)\,du = \frac{1}{2}a_0$, then show that

$$c_n = \frac{1}{2}(a_n - ib_n), \qquad c_{-n} = \frac{1}{2}(a_n + ib_n) \tag{3.9}$$

for all $n \in \mathbb{N}$. Finally, use these formulas to show that

$$c_n e^{2\pi inu} + c_{-n}e^{-2\pi inu} = a_n\cos(2\pi nu) + b_n\sin(2\pi nu)$$

for all $n \in \mathbb{N}$. Equations (3.7) and (3.8) embody the "classical Fourier series" expansion for real valued functions in $L^2[0,1]$.

In practice, it's often useful to rewrite eq. (3.7) in the form

$$f(u) = \frac{1}{2}a_0 + \sum_{n=1}^{\infty}A_n\cos[2\pi n(u - \theta_n)], \tag{3.10}$$

where $A_n$ satisfies

$$A_n^2 = a_n^2 + b_n^2 \tag{3.11}$$

and $\theta_n$ is any solution to

$$\cos(2\pi n\theta_n) = \frac{a_n}{A_n} \qquad \text{and} \qquad \sin(2\pi n\theta_n) = \frac{b_n}{A_n}. \tag{3.12}$$

The parameters $A_n$ and $\theta_n$ are not uniquely determined by eqs. (3.11) and (3.12), but in practice natural candidates for $A_n$ and $\theta_n$ often present themselves. We'll see an example of this shortly.

Equation (3.10) says that $f(u)$ may be written as a constant plus a series whose components are oscillatory functions of the form

$$A_n\cos[2\pi n(u - \theta_n)].$$

For each $n \in \mathbb{N}$, the parameter $|A_n|$ is the "amplitude" of the oscillation and $\theta_n$ is a "phase." As $u$ varies from 0 to 1, the function $\cos[2\pi n(u - \theta_n)]$ oscillates through exactly $n$ cycles. Hence, if $u$ represents *time* (measured, for example, in seconds), then $n$ may be interpreted as representing *frequency*, i.e., cycles per second.

If the function $f()$ is a pdf, $\frac{1}{2}a_0 = \int_0^1 f(u)du = 1$, so eq. (3.10) may be written

$$f(u) = 1 + \sum_{n=1}^{\infty} A_n \cos[2\pi n(u - \theta_n)]. \tag{3.13}$$

Hence, $f$ is the pdf of a $U[0,1)$ random variable iff $A_n = 0$ for all $n \in \mathbb{N}$. Furthermore, it follows from eq. (3.9) that

$$A_n^2 = a_n^2 + b_n^2 = 4c_n c_{-n} = 4|c_n|^2 \quad \Rightarrow \quad |A_n| = 2|c_n| \tag{3.14}$$

for all $n \in \mathbb{N}$. Hence we've proven the following important fact. **Proposition 3.1**. Assume that $f()$ is a pdf on $[0,1)$. Then $f()$ is the pdf of a $U[0,1)$ random variable iff $c_n = 0$ for all $n \in \mathbb{N}$.

We now turn to the topic of Fourier transforms. Fourier transforms of a function are defined for complex valued functions, but we are mainly interested in this document with finding the Fourier transforms of real valued functions. In particular, we are interested in the Fourier transforms of probability density functions. In what follows, I'll give the definitions and major properties of Fourier transforms, then describe the simplifications that follow from restricting these functions to being real valued.

Let $f: \mathbb{R} \to \mathbb{C}$ and suppose that $f \in L^2(\mathbb{R})$. Several different definitions of the "Fourier transform" of $f$ exist. The definition that is most useful for the purposes of this paper is

$$\widehat{f}(\xi) \equiv \int_{-\infty}^{\infty} \exp(-2\pi i \xi x) f(x)\, dx \tag{3.15}$$

where $\xi \in \mathbb{R}$. Hence $\widehat{f}: \mathbb{R} \to \mathbb{C}$. If $f$ is the pdf of a random variable $X$, a convenient way to remember eq. (3.15) is

$$\widehat{f}(\xi) = \mathbb{E}[\exp(-2\pi i \xi X)]. \tag{3.16}$$

Perhaps the most common alternative definition of the Fourier transform is

$$\widetilde{f}(\omega) \equiv \int_{-\infty}^{\infty} \exp(-i\omega x) f(x)\, dx \tag{3.17}$$

where $\omega \in \mathbb{R}$. For example, this is the way the mathematical software package "Maple" defines the Fourier transform. By the substitution $\omega = 2\pi\xi$, we see that $\widehat{f}(\xi) = \widetilde{f}(2\pi\xi)$.

If $f$ is the pdf of a random variable $X$, then there is one more transform of relevance. The "characteristic function" of $X$ is defined as

$$\varphi_X(\zeta) \equiv \mathbb{E}[\exp(i\zeta X)] = \int_{-\infty}^{\infty} \exp(i\zeta x) f(x)\, dx. \tag{3.18}$$

The substitution $\zeta = -2\pi\xi$ shows that $\widehat{f}(\xi) = \varphi_X(-2\pi\xi)$.

Both versions of Fourier transform and characteristic functions have some useful mathematical properties. In what follows I'll list some of these properties for $\widehat{f}(\xi)$, generally without proofs. My main references for this material are the Wikipedia article on Fourier transforms, and Feller's [5] chapter on characteristic functions.

The Fourier transform $\widehat{f}()$ of an integrable function $f()$ is uniformly continuous. Equation (3.15) implies that

$$\widehat{f}(0) = \int_{-\infty}^{\infty} f(x)\,dx. \tag{3.19}$$

If $f()$ is a pdf, eq. (3.19) shows that $\widehat{f}(0) = 1$. The Riemann-Lebesgue lemma states that $\widehat{f}(\xi) \to 0$ as $|\xi| \to \infty$.

Let $\mathcal{F}$ denote the operator that maps $f()$ into $\widehat{f}()$:

$$(\mathcal{F}f)(\xi) \equiv \widehat{f}(\xi).$$

**Inverses**. Under weak conditions, all of these transforms have inverses, so $f()$ may be retrieved from $\widehat{f}()$, $\tilde{f}()$, and $\varphi_X()$. For $\widehat{f}()$ the inverse is given by

$$f(x) = \left(\mathcal{F}^{-1}\widehat{f}\right)(x) = \int_{-\infty}^{\infty} \exp(2\pi i \xi x)\widehat{f}(\xi)\,d\xi. \tag{3.20}$$

**Duality**. Suppose that $f \in L^2(\mathbb{R})$ and let $g(\xi) \equiv \widehat{f}(\xi)$. Then eq. (3.20) implies that $\widehat{g}()$ exists and is given by $\widehat{g}(\xi) = f(-\xi)$. This is called the principle of duality.

**Linearity**. A fundamental property of $\mathcal{F}$ is that it's a *linear* operator. That is, for any real numbers $a$ and $b$, and for any two functions $f$ and $g$ in $L^2(\mathbb{R})$, $af + bg$ is also in $L^2(\mathbb{R})$ and

$$\mathcal{F}(af + bg)(\xi) = a\widehat{f}(\xi) + b\widehat{g}(\xi). \tag{3.21}$$

**Shift Property**. Define $g(x) \equiv f(x - a)$ where $a$ is a fixed real number. Then

$$\widehat{g}(\xi) = e^{-2\pi i a \xi}\widehat{f}(\xi). \tag{3.22}$$

**Scaling Property**. Define $h(x) \equiv f(bx)$ where $b \neq 0$. Then

$$\widehat{h}(\xi) = \frac{1}{|b|}\widehat{f}\left(\frac{\xi}{b}\right). \tag{3.23}$$

**Shift and Scale with Random Variables**. Let $X$ be a random variable, and let $Y \equiv cX + d$ where $c$ and $d$ are constants with $c > 0$. Then

$$\begin{aligned}
\varphi_Y(\zeta) &= \mathbb{E}[\exp(i\zeta Y)] = \mathbb{E}\{\exp[i\zeta(cX + d)]\} \\
&= \exp(id\zeta)\mathbb{E}\{\exp[i(c\zeta)X]\} = \exp(id\zeta)\varphi_X(c\zeta).
\end{aligned} \tag{3.24}$$

Let $f$ be the pdf of $X$ and $g$ be the pdf of $Y$. By an argument like that of eq. (3.24) (making use of the linearity of $\mathbb{E}()$) one may show that

$$\widehat{g}(\xi) = \exp(-2\pi i d\xi)\widehat{f}(c\xi). \tag{3.25}$$

This is a very useful fact.

Before moving on to the next topic, it's convenient to define *rectangular* and *triangular* functions. The *unit rectangular function* is defined as

$$\text{rect}(x) \equiv \begin{cases} 1 & \text{if } |x| < \frac{1}{2}, \\ 0 & \text{if } |x| > \frac{1}{2}, \\ \frac{1}{2} & \text{if } |x| = \frac{1}{2}. \end{cases} \tag{3.26}$$

The basic symmetric *triangular function* is defined as

$$\text{tri}(x) \equiv \max\{0, 1 - |x|\} = \begin{cases} 1 - |x| & \text{if } |x| < 1, \\ 0 & \text{if } |x| \geq 1. \end{cases} \tag{3.27}$$

**Convolutions**. Let $f$ and $g$ be elements of $L^2(\mathbb{R})$. The *convolution* of $f$ and $g$, written $f*g$, is the function defined by the following integral:

$$(f*g)(u) \equiv \int_{-\infty}^{\infty} f(x)g(u-x)\,dx. \tag{3.28}$$

Under these assumptions, $f*g$ is also an element of $L^2(\mathbb{R})$. For example, suppose that

$$f(x) = g(x) = \text{rect}(x).$$

The reader may confirm that

$$(f*g)(x) = \text{tri}(x).$$

Convolutions are important in probability theory because of the following fact: if $f$ is the pdf of $X$ and $g$ is the pdf of $Y$ and $X$ and $Y$ are independent random variables, then $f*g$ is the pdf of $X + Y$. For example, if both $X$ and $Y$ are uniformly distributed on $\left[-\frac{1}{2}, \frac{1}{2}\right]$, so $f(x) = g(x) = \text{rect}(x)$, and $X$ and $Y$ are independent, then the pdf of $X + Y$ is $\text{tri}(x)$.

**Convolution Theorem**: if $h = f*g$, then

$$\widehat{h}(\xi) = \widehat{f}(\xi)\widehat{g}(\xi). \tag{3.29}$$

In words, taking Fourier transforms converts the mathematically complicated operation of convolution into the simplier operation of multiplication. (This property holds for other transforms and for characteristic functions.)

Now suppose that $f$ is real valued. It follows from eq. (3.15) that $\widehat{f}(-\xi) = \overline{\widehat{f}(\xi)}$. Equation (3.19) implies that $\widehat{f}(0) \in \mathbb{R}$. If $f(x) \geq 0$ for all $x$, then $\left|\widehat{f}(\xi)\right| \leq \widehat{f}(0)$ for all $\xi$. Fourier transforms are integrals of a complex valued function, but it is easy to rewrite the definitions so that the integrals are integrals of real valued functions when $f$ is real valued. Specifically,

$$\widehat{f}(\xi) = u(\xi) + iv(\xi), \tag{3.30}$$

where

$$
\begin{aligned}
u(\xi) &= \int_{-\infty}^{\infty} \cos(-2\pi\xi x) f(x)\,dx = \int_{-\infty}^{\infty} \cos(2\pi\xi x) f(x)\,dx, \\
v(\xi) &= \int_{-\infty}^{\infty} \sin(-2\pi\xi x) f(x)\,dx = -\int_{-\infty}^{\infty} \sin(2\pi\xi x) f(x)\,dx.
\end{aligned}
\tag{3.31}
$$

Note that $u()$ is even and $v()$ is odd. The two integrals that appear in eq. (3.31),

$$
\int_{-\infty}^{\infty} \sin(2\pi\xi x) f(x)\,dx \qquad \text{and} \qquad \int_{-\infty}^{\infty} \cos(2\pi\xi x) f(x)\,dx
$$

are known as the "Fourier sine" and "Fourier cosine" transformations, respectively. If $f()$ is the pdf of a random variable $X$, then $u(\xi)$ and $v(\xi)$ may be rewritten in terms of expected values of real random variables, as follows.

$$
\begin{aligned}
u(\xi) &= \mathbb{E}[\cos(-2\pi\xi X)] = \mathbb{E}[\cos(2\pi\xi X)], \\
v(\xi) &= \mathbb{E}[\sin(-2\pi\xi X)] = -\mathbb{E}[\sin(2\pi\xi X)].
\end{aligned}
\tag{3.32}
$$

If $f()$ is even, then $\widehat{f}() = u()$ is real valued and even. Hence, if $f()$ is the pdf of $X$, then $\widehat{f}()$ is real valued and even if $X$ is symmetrically distributed around $X = 0$.

Appendix B contains a small table of Fourier transforms of probability density functions adapted from a table in Feller [5], page 503.

As a final topic in this section, it is interesting to "compare and contrast" Fourier transforms and Fourier series. The following table shows the similarities and differences.

|  | Transform | Series |
|---|---|---|
| Domain | $f \in L^2(\mathbb{R})$ | $f \in L^2[0,1]$ |
| Co-domain | $\widehat{f} \in L^2(\mathbb{R})$ | $(c_n \colon n \in \mathbb{Z}) \in \ell_2(\mathbb{Z})$ |
| $\mathcal{F}$ | $\widehat{f}(\xi) = \int_{-\infty}^{\infty} e^{-2\pi i \xi x} f(x)\,dx$ | $c_n = \int_0^1 e^{-2\pi i n x} f(x)\,dx$ |
| $\mathcal{F}^{-1}$ | $f(x) = \int_{-\infty}^{\infty} \widehat{f}(\xi) e^{2\pi i \xi x}\,d\xi$ | $f(x) = \sum_{n \in \mathbb{Z}} c_n e^{2\pi i n x}$ |

In both transforms and series, the operator $\mathcal{F}$ maps a function in one domain into a function in a co-domain. These maps are one-to-one, and the inverse map $\mathcal{F}^{-1}$ shows how to retrieve the original function.

## 4. Benford Analysis Example 1: Lognormal Random Variables

We now resume our study of the Benford analysis of a positive random variable $X$ given a base $b > 1$. Recall the following standard notations: first, define $Y \equiv \log_b(X) = \Lambda_b \ln(X)$. Next, let

$$G() \text{ and } g() \quad \text{denote the cdf and pdf of } Y,$$
$$\widetilde{G}() \text{ and } \widetilde{g}() \quad \text{denote the cdf and pdf of } \langle Y \rangle.$$

The Fourier series representation of $\widetilde{g}(u)$ is

$$\widetilde{g}(u) = \sum_{n \in \mathbb{Z}} c_n e^{2\pi i n u} \tag{2.14}$$

where the Fourier coefficients $c_n$ for all $n \in \mathbb{Z}$ are given by

$$c_n = \widehat{g}(n). \tag{2.16}$$

As $\widetilde{g}()$ is a pdf, eq. (2.14) may be rewritten in the form

$$\widetilde{g}(u) = 1 + \sum_{n=1}^{\infty} A_n \cos[2\pi n(u - \theta_n)], \tag{4.1}$$

so

$$\|\widetilde{g} - 1\|_{\infty} = \left\| \sum_{n=1}^{\infty} A_n \cos[2\pi n(u - \theta_n)] \right\|_{\infty} \leq \sum_{n=1}^{\infty} |A_n| = 2 \sum_{n=1}^{\infty} |c_n|. \tag{4.2}$$

Our first illustration of the use and utility of these constructions is the analysis of lognormal random variables.

Suppose that $X \sim \text{Lognormal}(\mu, \sigma^2)$. As noted in the paragraph following eq. (2.13), it follows that $Y \sim N\big(\Lambda_b \mu, (\Lambda_b \sigma)^2\big)$. Hence, to proceed we need to find $\widehat{g}()$ for random variables of this type. Let $Z$ denote a standard normal random variable and let $\phi()$ denote the pdf of $Z$. Let $Y \equiv sZ + m$ where $s$ and $m$ are constants with $s > 0$, so $Y \sim N(m, s^2)$. Let $g()$ denote $Y$'s pdf. From the first entry in the table of Fourier transforms (Appendix B), $\widehat{\phi}(\xi) = \exp(-2\pi^2 \xi^2)$, and from eq. (3.25) it follows that

$$\widehat{g}(\xi) = \exp(-2\pi i m \xi) \exp\big(-2\pi^2 s^2 \xi^2\big).$$

Putting these facts together,

$$\widehat{g}(\xi) = \exp(-2\pi i \Lambda_b \mu \xi) \exp\big(-2\pi^2 (\Lambda_b \sigma)^2 \xi^2\big),$$

so

$$|c_n| = \exp\big(-2\pi^2 (\Lambda_b \sigma)^2 n^2\big) = R(\Lambda_b \sigma)^{n^2} > 0 \quad \text{for all} \quad n \in \mathbb{N}, \tag{4.3}$$

where the function $R()$ is defined as

$$R(s) \equiv \exp\big(-2\pi^2 s^2\big). \tag{4.4}$$

Hence,

$$\|\widetilde{g} - 1\|_\infty \leq 2\sum_{n=1}^{\infty} R(\Lambda_b \sigma)^{n^2} \leq \frac{2R(\Lambda_b \sigma)}{1 - R(\Lambda_b \sigma)}. \tag{4.5}$$

From eq. (4.3) we see that $X$ is not $b$-Benford. However, $\|\widetilde{g} - 1\|_\infty \to 0$ as $\sigma \to \infty$, so $X$ is asymptotically $b$-Benford. In fact, $\|\widetilde{g} - 1\|_\infty$ becomes quite small for even moderate values of $\sigma$, and it's fair to say in these cases that $X$ is "effectively" $b$-Benford. To see this, it's sufficient to note how quickly $R(s)$ decreases to zero as $s$ increases. When $s \geq 1$ we have $R(s) \leq 2.6753 \times 10^{-9}$, and the upper bound on $\|\widetilde{g} - 1\|_\infty$ given in eq. (4.5) is less than or equal to $5.3506 \times 10^{-9}$.

## 5. Base Dependence of Benford Random Variables.

My grandfather considered his "law of anomalous numbers" as evidence of a "real world" phenomenon. He realized that geometric sequences and exponential functions[6] are generally base 10 Benford, and on this basis he wrote (Benford, (1938) [1]):

> If the view is accepted that phenomena fall into geometric series, then it follows that the observed logarithmic relationship is not a result of the particular numerical system, with its base, 10, that we have elected to use. Any other base, such as 8, or 12, or 20, to select some of the numbers that have been suggested at various times, would lead to similar relationships; for the logarithmic scales of the new numerical system would be covered by equally spaced steps by the march of natural events. As has been pointed out before, the theory of anomalous numbers is really the theory of phenomena and events, and the *numbers* but play the poor part of lifeless symbols for living things.

This argument seems compelling, and it might seem to apply to Benford random variables as well as to geometric sequences and exponential functions. It is therefore somewhat surprising to observe that a random variable that is base $b$ Benford is not generally base $c$ Benford when $c \neq b$. In this section and the next I examine some of the issues of "base dependence" of Benford random variables.

We begin with a definition: following Wójcik [11], the "**Benford spectrum**" of a positive random variable $X$, denoted $B_X$, is defined as

$$B_X \equiv \{b \in (1, \infty) \colon X \text{ is } b\text{-Benford}\}. \tag{5.1}$$

The Benford spectrum of $X$ may be empty. For example, as shown above, $B_X = \emptyset$ if $X$ is a lognormal random variable. In fact, the Benford spectra of essentially all of the usual families of random variables used in statistics (e.g., Gamma, F, Weibull) are empty.

There are a couple of elementary propositions about $B_X$ we can prove at this time.

---

[6] See "Benford's Law, A Growth Industry" by Kenneth Ross [6] for further information on this topic.

**Proposition 5.1**. The set $B_X$ is bounded above. **Proof**. If $X$ is a positive random variable, then $\log_b(X) = \Lambda_b \ln(X)$. Let $g_b()$ and $g_e()$ denote the pdfs of $\log_b(X)$ and $\ln(X)$, respectively. Then eq. (3.25) implies that

$$\widehat{g}_b(\xi) = \widehat{g}_e(\Lambda_b \xi) \quad \text{for all } \xi \in \mathbb{R}.$$

Now suppose that $X$ is $b$-Benford. It follows from Proposition 3.1 that $\widehat{g}_b(1) = \widehat{g}_e(\Lambda_b) = 0$. As $\widehat{g}_e()$ is the Fourier transform of a pdf, it follows that $\widehat{g}_e(0) = 1$ and $\widehat{g}_e()$ is continuous. Hence, there exists $r > 0$ such that $\widehat{g}_e(w) > 0$ for all $w \in [0, r]$. Therefore

$$\Lambda_b = \frac{1}{\ln(b)} > r \quad \Rightarrow \quad b < e^{1/r}.$$

My proof of the next proposition makes use of the following fact (**Proposition 4.3 (iii), page 44, of Berger and Hill [4]**): a random variable $Y$ is u. d. mod 1 if and only if $nY + r$ is u. d. mod 1 for every integer $n \neq 0$ and $r \in \mathbb{R}$.

**Proposition 5.2**. If $b \in B_X$, then $\sqrt[m]{b} = b^{1/m} \in B_X$ for every $m \in \mathbb{N}$. **Proof**. As $X$ is $b$-Benford,

$$X = b^Y \quad \text{and} \quad \langle Y \rangle \sim U[0, 1).$$

Hence, for any $m \in \mathbb{N}$,

$$X = \left(b^{1/m}\right)^{mY} \quad \text{and} \quad \langle mY \rangle \sim U[0, 1)$$

from Berger and Hill's Proposition 4.3. As $b^{1/m} > 1$, it follows that $b^{1/m} \in B_X$.

**Bibliographic Notes**. Wójcik [11] attributes Proposition 5.1 to Schatte (1981) and rediscovered by Whittaker (1983). He attributes Proposition 5.2 to Whittaker.

It's appropriate at this point to insert a pair of propositions that are related to Berger and Hill's Proposition 4.3.

**Proposition 5.3**. If $X$ is $b$-Benford, then $cX$ is $b$-Benford for any constant $c > 0$. **Proof**. As $X$ is $b$-Benford, $\log_b(X)$ is u. d. mod 1. But $\log_b(cX) = \log_b(X) + \log_b(c)$, so $\log_b(cX)$ is u. d. mod 1 from Berger and Hill's Proposition 4.3. Hence $cX$ is $b$-Benford.

We say of this result that Benford random variables are "scale-invariant." This is only to be expected if Benford's Law holds for empirical data. For example, one of the variables included in grandfather's data set was "area of rivers." He doesn't state the *unit* in which area is measured: it could be acres, or hectares, or something else altogether. Proposition 5.3 says that the chosen units don't affect the "Benfordness" of the underlying random variable $X$.

**Proposition 5.4**. Let $X$ and $Y$ denote independent positive random variables. If $X$ is $b$-Benford, then the product $XY$ is $b$-Benford. **Proof**. Let $g_x()$, $g_y()$, and $g_{xy}()$ denote the pdfs of $\log_b(X)$, $\log_b(Y)$, and $\log_b(XY)$, respectively. As $X$ and $Y$ are independent and $\log_b(XY) = \log_b(X) + \log_b(Y)$, it follows that $g_{xy} = g_x * g_y$. From the convolution theorem, it follows that $\widehat{g}_{xy}(\xi) = \widehat{g}_x(\xi)\widehat{g}_y(\xi)$ for all $\xi \in \mathbb{R}$. As $X$ is $b$-Benford, $\widehat{g}_x(n) = 0$ for all $n \in \mathbb{N}$. Hence $\widehat{g}_{xy}(n) = 0$ for all $n \in \mathbb{N}$, and it follows that $XY$ is $b$-Benford.

Proposition 5.4 has an immediate corollary. **Proposition 5.5**. If $X$ and $Y$ are independent positive random variables, then $B_X \cup B_Y \subseteq B_{XY}$.

Suppose that $X$ is $b$-Benford. It is of some interest to examine how the distribution of $\langle \log_c(X) \rangle$ varies as a function of $c$. Let $g_c()$ denote the pdf of $\log_c(X)$ and let $\widetilde{g}_c()$ denote the pdf of $\langle \log_c(X) \rangle$. We know from eq. (4.1) that $\widetilde{g}_c()$ may be written in the form

$$\widetilde{g}_c(u) = 1 + \sum_{n=1}^{\infty} A_n \cos[2\pi n(u - \theta_n)],$$

so we want to examine how each $A_n$ and $\theta_n$ varies as a function of $c$ (and of any parameters that govern the distribution of $X$). I carried out such a program and reported my results in my 2017 paper [3]. Appendix A of this paper contains a sketch of my methods. Here I will briefly summarize some of my results. First, it may be shown that the dependence of $A_n$ and $\theta_n$ on $c$ is entirely conveyed through a parameter $\rho$ defined as

$$\rho \equiv \frac{\ln(b)}{\ln(c)}. \tag{5.2}$$

Let $H()$ be a "seed function" as described in the appendix. Every cdf is a legitimate seed function, and I used the cdfs of standard families of random variables as seed functions to generate five families of $b$-Benford random variables (which explains the names I gave to these families). For any differentiable seed function $H()$, I let $h(y) \equiv H'(y)$. The details of the specification of three of these families are shown in the following table.

| Family name | Parameters | $h(y) \equiv H'(y)$ | Notes |
|---|---|---|---|
| Gauss-Benford | $\mu, \sigma$ | $\frac{1}{\sigma}\phi\left(\frac{y-\mu}{\sigma}\right)$ | $\phi()$ is the standard normal pdf |
| Cauchy-Benford | $\mu, \sigma$ | $\frac{1}{\pi\sigma}\left[1 + \left(\frac{y-\mu}{\sigma}\right)^2\right]^{-1}$ | |
| Laplace-Benford | $\mu, \sigma$ | $\frac{1}{2\sigma}\exp\left(-\frac{|y-\mu|}{\sigma}\right)$ | |

The function $h()$ for each of these families is symmetrically distributed around a point $\mu$, and this fact yields some important analytic simplifications. First, this symmetry implies that $\theta_n$ does not depend on $n$ but is given by

$$\theta_n = \rho\left(\frac{1}{2} + \mu\right) \tag{5.3}$$

for all $n \in \mathbb{N}$. Second, this symmetry implies that the $A_n$'s all have the form

$$A_n = \frac{2\sin(\pi\rho n)}{\pi\rho n}C_n \tag{5.4}$$

where $C_n$ varies between the families. The dependence of $C_n$ on $\rho$ and the parameters for these three families is shown in the following table.

| Family | $C_n$ |
|---|---|
| Gauss-Benford | $\exp(-2\pi^2\sigma^2\rho^2 n^2)$ |
| Cauchy-Benford | $\exp(-2\pi\sigma\rho n)$ |
| Laplace-Benford | $(1 + 4\pi^2\sigma^2\rho^2 n^2)^{-1}$ |

Note that $C_n > 0$ for all of these families. Hence, $A_n = 0$ iff $\sin(\pi n\rho) = 0$, and this can happen for all $n \in \mathbb{N}$ iff $\rho$ is a positive integer, say $m$. But

$$\rho = \frac{\ln(b)}{\ln(c)} = m \quad \Leftrightarrow \quad c = b^{1/m}. \tag{5.5}$$

Hence, $B_X$ for all these random variables is precisely the set of integral roots of $b$. Also, note that the parameter $\mu$ affects only the phase, while the parameter $\sigma$ affects only $C_n$. In somewhat more detail, $C_n$ is a decreasing function of $\sigma$. (Also note that $C_n$ is a decreasing function of both $n$ and $\rho$.)

## 6. Whittaker's Random Variable.

All of the non-empty Benford spectra $B_X$ constructed above are discrete. The question naturally arises: is it possible for a random variable to have a Benford spectrum that consists, at least in part, of intervals of positive length? Whittaker [10] showed by an example that such a random variable exists.

Let $b > 1$ and let

$$g(y) \equiv \frac{1 - \cos(2\pi y)}{2\pi^2 y^2}. \tag{6.1}$$

A plot of $g(y)$ between $-3$ and $+3$ is shown below. It may be shown that $\int_{-\infty}^{\infty} g(y)\, dt = 1$, so $g()$ is a legitimate pdf. Suppose that $Y \sim g()$ and $X \equiv b^Y$, so

$$Y = \log_b(X) = \frac{\ln(X)}{\ln(b)}. \tag{6.2}$$

This $X$ is *Whittaker's random variable*. From entry 5 of the table of Fourier transforms in Appendix B we see that

$$\widehat{g}(\xi) = \max(0, 1 - |\xi|). \tag{6.3}$$

As $\widehat{g}(n) = 0$ for all integers $n \neq 0$, it follows that $\langle Y \rangle \sim U[0, 1)$, and hence that $X$ is $b$-Benford.

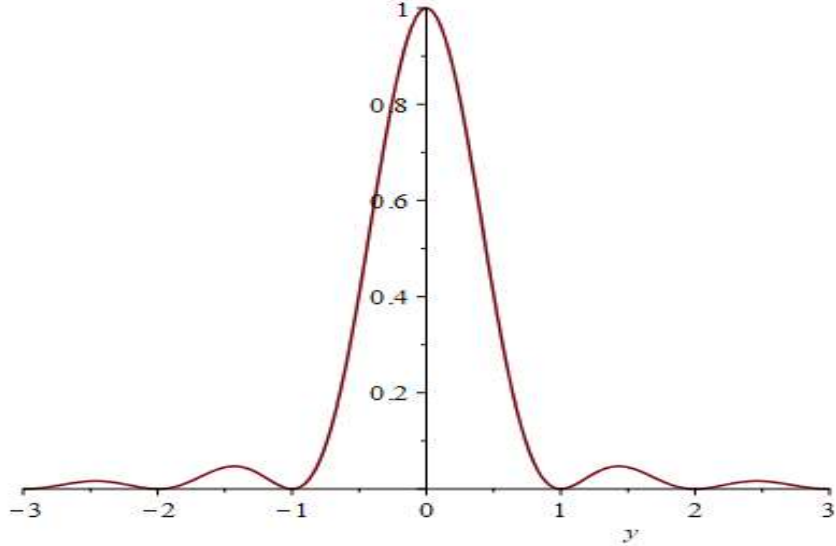

Now let $1 < c < b$ and define

$$Y_c \equiv \log_c(X) = \frac{\ln(X)}{\ln(c)} = \frac{\ln(b)}{\ln(c)} \cdot \frac{\ln(X)}{\ln(b)} = \rho Y. \qquad (6.4)$$

Let $g_c()$ denote the pdf of $Y_c$. It follows from eq. (3.25) that

$$\widehat{g}_c(\xi) = \widehat{g}(\rho\xi) = \max(0, 1 - \rho|\xi|).$$

As $\widehat{g}_c(\xi) = 0$ whenever $\rho|\xi| \geq 1$ and $\rho > 1$, it follows that $\widehat{g}_c(n) = 0$ for all integers $n \neq 0$. Therefore, $\langle Y_c \rangle \sim U[0, 1)$, and hence $X$ is $c$-Benford. Hence, Whittaker's random variable has the remarkable feature that it is $c$-Benford for all $1 < c \leq b$. Hence $B_X = (1, b]$.

Let's step back for a minute and consider what goes into the construction of a random variable with the properties of Whittaker's random variable. The crucial goal is to find a pdf $g()$ whose Fourier transform $\widehat{g}()$ has bounded support. A reasonable tactic to find $g()$ is to start with a pdf $f()$ with bounded support. If the Fourier transform $\widehat{f}()$ can be rescaled into a legitimate pdf, say $\gamma()$, then by duality the Fourier transform $\widehat{\gamma}()$ is proportional to $f()$ and therefore has bounded support. What are the requirements on $f()$ that allow $\widehat{f}()$ to be "rescaled into a legitimate pdf"? First, $\widehat{f}()$ has to be real-valued, which requires that $f()$ be an even function. Second, $\widehat{f}(\xi)$ has to be non-negative for all $\xi \in \mathbb{R}$. One way to satisfy the second requirement is to require $f()$ to be the convolution of some even function $\psi()$ with itself, as the convolution theorem then guarantees that $\widehat{f}(\xi) = \widehat{\psi}(\xi)^2$. The function $g()$ of Whittaker's random variable was constructed along these lines with $\psi()$ equal to rect(). This is probably the simplest way, but not the only way, to construct a random variable with the required properties. I think Whittaker's whole construction is remarkably clever.

**Appendix A: Seed Functions**

The first purpose of this appendix is to describe the use of seed functions to generate random variables that are Benford relative to a given base. Let $X$ be a $b$-Benford random variable generated by a seed function $H$. A second purpose is to sketch how seed functions may be used to carry out a Benford analysis on $\log_c(X)$ where $c \neq b$.

**Definition.** A function $H\colon \mathbb{R} \to \mathbb{R}$ is a seed function if it satisfies the following three conditions.

$$(1) \quad \lim_{y \to -\infty} H(y) = 0;$$

$$(2) \quad \lim_{y \to +\infty} H(y) = 1;$$

$$(3) \quad H(y) - H(y-1) \geq 0 \quad \forall y \in \mathbb{R}.$$

We say of the last condition that $H$ is "unit interval increasing." If $H$ is increasing, then $H$ is unit interval increasing, but the converse is not always true. Hence every cdf is a seed function, but not every seed function is a cdf.

Let $H$ be a seed function and define

$$g(y) \equiv H(y) - H(y-1) \quad \forall y \in \mathbb{R}. \tag{A.1}$$

The assumption that $H$ is unit interval increasing guarantees that $g(y) \geq 0$ for all $y \in \mathbb{R}$. Also, it may be shown that

$$\int_{-\infty}^{\infty} g(y)\, dy = 1. \tag{A.2}$$

This is geometrically evident if $H$ is an increasing function, as the graph of the function $H(y-1)$ is just the graph of the function $H(y)$ shifted to the right by 1. The integral of eq. (A.2) gives the area between the two graphs as the integral of the *vertical* distance between the two curves. But the area between the two curves is also equal to the integral of the *horizontal* distance between the two curves, which is just $\int_0^1 1\, dv = 1$.

Hence, $g()$ may be regarded as the pdf of a random variable $Y$.

**Proposition A.1**: If $Y \sim g()$, then $\langle Y \rangle \sim U[0,1)$. **Proof.** For any $N \in \mathbb{N}$ and $u \in [0,1)$, consider

$$\sum_{k=-N}^{N} g(k+u) = \sum_{k=-N}^{N} [H(k+u) - H(k-1+u)]. \tag{A.3}$$

This sum telescopes, so

$$\sum_{k=-N}^{N} g(k+u) = H(N+u) - H(-N-1+u).$$

Letting $N \to \infty$, we find that

$$\sum_{k=-\infty}^{\infty} g(k+u) = 1$$

for every $u \in [0, 1)$. As $\sum_{k \in \mathbb{Z}} g(k+u)$ is just the pdf of $\langle Y \rangle$, this proves the proposition.

**Corollary**. Let $b > 1$. If $Y \sim g()$, then $X \equiv b^Y$ is $b$-Benford.

Now let $g()$, $Y$, $b$, and $X$ be defined as above, and let $c > 1$. We are interested in the distribution of $\log_c(X)$ and $\langle \log_c(X) \rangle$. Let $Y_c \equiv \log_c(X)$, let $g_c()$ denote the pdf of $Y_c$, and let $\widetilde{g}_c()$ denote the pdf of $\langle \log_c(X) \rangle$. We know that the Fourier series expansion of $\widetilde{g}_c()$ is

$$\widetilde{g}_c(u) = \sum_{n \in \mathbb{Z}} \widehat{g}_c(n) e^{2\pi i n u}. \tag{A.4}$$

From eq. (6.4) we know that $Y_c = \rho Y$, and it follows from eq. (3.25) that

$$\widehat{g}_c(\xi) = \widehat{g}(\rho \xi). \tag{A.5}$$

Hence, we may find the Fourier series expansion of $\widetilde{g}_c()$ once we've found $\widehat{g}()$, defined as

$$\widehat{g}(\xi) \equiv \int_{-\infty}^{\infty} e^{-2\pi i \xi y} g(y) \, dy = \int_{-\infty}^{\infty} e^{-2\pi i \xi y} \left[ H(y) - H(y-1) \right] dy.$$

Note first that $\widehat{g}_c(0) = \widehat{g}(0) = \int_{\mathbb{R}} g(y) \, dy = 1$. To evaluate $\widehat{g}(\xi)$ when $\xi \neq 0$, assume that $H()$ is differentiable and let $h(y) \equiv H'(y)$. Then integration by parts yields

$$\widehat{g}(\xi) = \frac{1}{2\pi i \xi} \left( 1 - e^{-2\pi i \xi} \right) \widehat{h}(\xi)$$

$$= \frac{e^{-\pi i \xi}}{2\pi i \xi} \left( e^{\pi i \xi} - e^{-\pi i \xi} \right) \widehat{h}(\xi) = \frac{e^{-\pi i \xi}}{\pi \xi} \sin(\pi \xi) \widehat{h}(\xi)$$

Hence, the Fourier coefficient $\widehat{g}_c(n)$ for any integer $n \neq 0$ is given by

$$\widehat{g}_c(n) = \widehat{g}(\rho n) = \frac{e^{-i\pi \rho n}}{\pi \rho n} \sin(\pi \rho n) \widehat{h}(\rho n). \tag{A.6}$$

Equation (A.6) is used to derive eqs. (5.3) and (5.4) as described in my 2017 paper. But we may observe right away from eq. (A.6) that $\widehat{g}_c(n) = 0$ for all $n \neq 0$ iff either (1) $\widehat{h}(\rho) = 0$, or (2) $\rho$ is an integer. As noted above, $\rho$ is an integer iff $c = \sqrt[m]{b}$ for some $m \in \mathbb{N}$.

**Appendix B: A Small Table of Fourier Transforms**

Feller [5] gives a table of characteristic functions of selected probability functions. I've adapted his table to give the Fourier transforms of 8 of his 10 densities.

| No. | Name | Density $f(x)$ | Interval | Fourier Transform $\widehat{f}(\xi)$ |
|-----|------|----------------|----------|--------------------------------------|
| 1 | $N(0,1)$ | $(2\pi)^{-1/2}e^{-x^2/2}$ | $\mathbb{R}$ | $\exp(-2\pi^2\xi^2)$ |
| 2 | $U[-a,a]$ | $1/2a$ | $[-a,a]$ | $\frac{\sin(2\pi a\xi)}{2\pi a\xi}$ |
| 3 | $U[0,a]$ | $1/a$ | $[0,a]$ | $\frac{1-\exp(-2\pi i a\xi)}{2\pi i a\xi}$ |
| 4 | Triangular | $\frac{1}{a}\left(1-\frac{|x|}{a}\right)$ | $|x|\leq a$ | $\frac{1-\cos(2\pi a\xi)}{2\pi^2 a^2\xi^2}$ |
| 5 | Dual of 4 | $\frac{1-\cos(2\pi ax)}{2a\pi^2 x^2}$ | $\mathbb{R}$ | $\max\left(0, 1-\frac{|\xi|}{a}\right)$ |
| 6 | $\Gamma(\alpha,1)$ | $\frac{1}{\Gamma(\alpha)}x^{\alpha-1}e^{-x}$ | $x>0$ | $(1+2\pi i\xi)^{-\alpha}$ |
| 7 | Bi. Exp. | $\frac{1}{2}e^{-|x|}$ | $\mathbb{R}$ | $\frac{1}{1+4\pi^2\xi^2}$ |
| 8 | Cauchy | $\frac{1}{\pi}\frac{a}{a^2+x^2}$ | $\mathbb{R}$ | $e^{-2\pi a|\xi|}$ |

**Notes on this table**.

(1) This rule has been summarized as "the Fourier transform of a Gaussian is a Gaussian."

(2) If $a=1$, then $\widehat{f}(0)=1$ and

$$\widehat{f}(\xi) = \frac{1}{2}\int_{-1}^{1}e^{-2\pi i\xi x}\,dx = -\frac{1}{4\pi i\xi}\left[e^{-2\pi i\xi x}\right]_{-1}^{1} = \frac{e^{2\pi i\xi}-e^{-2\pi i\xi}}{4\pi i\xi}$$
$$= \frac{2i\sin(2\pi\xi)}{4\pi i\xi} = \frac{\sin(2\pi\xi)}{2\pi\xi} \tag{B.1}$$

when $\xi\neq 0$. Now apply eq. (3.25) with $c=a$ and $d=0$ to get rule (2). Letting $a=\frac{1}{2}$ we obtain

$$\widehat{\text{rect}}(\xi) = \frac{\sin(\pi\xi)}{\pi\xi}. \tag{B.2}$$

(3) Suppose that $X\sim U[-a,a]$ and $Y\equiv\frac{1}{2}X+\frac{1}{2}a$, so $Y\sim U[0,a]$. Let $g()$ denote the pdf of $Y$. Recall that $\sin(\theta)=(2i)^{-1}\left(e^{i\theta}-e^{-i\theta}\right)$ and apply eq. (3.25) to the Fourier transform of rule 2 to yield

$$\widehat{g}(\xi) = \frac{1-e^{-2\pi i a\xi}}{2\pi i a\xi}.$$

(4) As $\text{tri}(x)=(\text{rect}*\text{rect})(x)$, it follows from the convolution theorem that

$$\widehat{\text{tri}}(\xi) = \widehat{\text{rect}}(\xi)^2 = \frac{\sin(\pi\xi)^2}{(\pi\xi)^2} = \frac{1-\cos(2\pi\xi)}{2(\pi\xi)^2}. \tag{B.3}$$

(5) The integral of $\widehat{\mathrm{tri}}(a\xi)$ over $\xi \in \mathbb{R}$ equals $1/a$, so $a \cdot \widehat{\mathrm{tri}}(a\xi)$ is a pdf. Note that $\widehat{f}(\xi) = 0$ whenever $|\xi| \geq a$. This is important. Feller notes of this result

> "This formula is of great importance because many Fourier-analytic proofs depend on the use of a characteristic function vanishing outside a finite interval."

(6) Let $X$ be a $\Gamma(\alpha, 1)$ random variable and $f()$ denote $X$'s pdf, shown in column 2. Let $\beta > 0$, define $Y \equiv \beta X$, and let $g()$ be $Y$'s pdf. Then $Y \sim \Gamma(\alpha, \beta)$ and

$$\widehat{g}(\xi) = \widehat{f}(\beta\xi) = (1 + 2\pi i\beta\xi)^{-\alpha}. \tag{B.4}$$

(7) The notation "Bi. Exp." is an abbreviation for "Bilateral Exponential." This pdf is also known as that of a Laplace$(0, 1)$ random variable.

(8) The Cauchy and Bilateral Exponential distributions are dual to one another.

Let $X$ be a random variable with pdf $f$ given by rule (1), (7), or (8) in this table, let $Y \equiv sX + m$ where $s > 0$, and let $g$ be $Y$'s pdf. Then $Y$ is distributed symmetrically around $m$ with a "spread" measured[7] by $s$, and eq. (3.25) with $c = s$ and $d = m$ may be used to find $\widehat{g}(\xi)$.

---

[7] For rule (1) $s$ is the standard deviation of $Y$. This is *not* true for the other two distributions.